\documentclass[aps,twocolumn,showpacs,groupedaddress,superscriptaddress]{revtex4}
\usepackage{CJK}
\usepackage{epsfig}
\usepackage{amstext}
\usepackage{amsmath}
\usepackage{amssymb}
\usepackage{graphicx}
\usepackage{latexsym}

\newcommand{\ket}[1]{|#1\rangle}

\newcommand{\sz}{\sigma^z}
\newcommand{\sx}{\sigma^x}
\newcommand{\sy}{\sigma^y}
\newcommand{\sigmam}{\sigma^-}
\newcommand{\sigmap}{\sigma^+}
\newcommand{\ext}[2]{|#1\rangle\langle#2|}   

\newcommand*{\dbar}{\mkern 3mu\mathchar '26\mkern -12mu d}

\newcommand{\PRA}[3] {Phys. Rev. A {\bf #1}, #2(#3)}

\newcommand{\PRE}[3] {Phys. Rev. E {\bf #1}, #2(#3)}
\newcommand{\PRL}[3] {Phys. Rev. Lett. {\bf #1}, #2(#3)}
\newcommand{\JPA}[3] {J. Phys. A {\bf #1}, #2 (#3)}

\newcommand{\EPJD}[3] {Eur. Phys. J. D {\bf #1}, #2 (#3)}

\newcommand{\JCP}[3] {J. Chem. Phys. {\bf #1}, #2(#3)}
\newcommand{\Science}[3] {Science {\bf #1}, #2(#3)}
\newcommand{\JAP}[3] {J. Appl. Phys. {\bf #1}, #2(#3)}

\def\dbarit {{\mathchar'26\mkern-11mud}}

\begin{document}
\begin{CJK*}{GB}{gbsn}
\title{Quantum Brayton cycle with coupled systems as working substance}
\author{X. L. Huang(黄晓理)}
\email{huangxiaoli1982@foxmail.com}
\affiliation{School of physics and electronic technology, \\
Liaoning Normal University, Dalian, 116029, China}
\author{L. C. Wang(王林成)}
\affiliation{School of physics and optoelectronic technology,\\
Dalian University of Technology, Dalian 116024 China}
\author{X. X. Yi(衣学喜)}
\email{yixx@dlut.edu.cn}
\affiliation{School of physics and optoelectronic technology,\\
Dalian University of Technology, Dalian 116024 China}

\date{\today}

\begin{abstract}
We explore the quantum version of Brayton cycle with a composite
system as the working substance. The actual Brayton cycle consists
of two adiabatic  and two isobaric processes.  Two pressures can be
defined in our isobaric process, one corresponds to the external
magnetic field (characterized by $F_x$) exerted on the system, while
the other corresponds  to the coupling constant between the
subsystems  (characterized by $F_y$). As a consequence, we can
define two types of quantum Brayton cycle for the composite system.
We find that the subsystem experiences a quantum Brayton cycle in
one quantum Brayton cycle (characterized by $F_x$), whereas the
subsystem's cycle is of quantum Otto in another Brayton cycle
(characterized by $F_y$). The efficiency for the composite system
equals to that for the subsystem in both cases, but the work done by
the total system are usually larger than the sum of work done by the
two subsystems. The other interesting finding is that for the cycle
characterized by $F_y$, the subsystem can be a refrigerator while
the total system is a heat engine. The result in the paper can be
generalized to a quantum Brayton cycle with a general coupled system
as the working substance.

\end{abstract}

 \maketitle
\section{Introduction}
It is well know that there are four basic thermodynamical processes
in classical thermodynamics: adiabatic process, isothermal process,
isochoric process and isobaric process. These four processes
correspond to   entropy, temperature, volume  and pressure being
kept unchanged, respectively. The study on the quantum version for
these processes can be dated back to the quantum adiabatic theorem
\cite{Born,Messiah}, it brings the adiabat to quantum region and
opens the door to quantum thermal dynamics. Recently, the isothermal
and isochoric processes are generalized  to quantum case
\cite{Kieu2004PRL,Kieu2006EPJD}, and the quantum Carnot cycle and
quantum Otto cycle are  discussed in \cite{Quan2007PRE,Quan2005PRE}.
With the definition of  quantum isobaric process\cite{Quan2009PRE},
almost  all  thermodynamical cycles, in particular Brayton cycle and
Diesel cycle \cite{Callenbook,Perrotbook},  are extended from
classical to quantum region.

The studies in the field of  quantum thermodynamics
\cite{Fialko2012PRL,He2002PRE,FeldmannPRE,Wu2006JCP,Lin2003PRE,Li2007JPA}
are usually focused on whether it can surpass the classical limit on
the efficiency and work extraction  in  a cycle
\cite{Kieu2004PRL,Kieu2006EPJD,Scully2003Science},  and how to
better the work extraction in a cycle \cite{WangPRE}. For example,
it is reported that one can extract work from a single heat bath via
vanishing quantum coherence \cite{Scully2003Science} and the
efficiency of a quantum heat engine can be higher than the classical
one due to the effects of squeezing  heat bath \cite{Huangsubmit}.
These studies can better our understanding of the fundamental
concept in thermodynamics and quantum mechanics, and bring new
insights into basic problems in quantum mechanics and thermodynamics
\cite{Quan2006,ScullyPRL}.

Coupled quantum system as working substance becomes an active topic
recently\cite{Zhang2007PRA, Wang2009PRE, Zhang2008EPJD,
Thomas2011PRE}. The reason is twofold.  First,  the entanglement is
one of the features that distinguish quantum and classical worlds,
the effects of quantum entanglement on the basic thermodynamical
quantities  are then attractive
\cite{Zhang2007PRA,Wang2009PRE,Zhang2008EPJD}. Second, for a cycle
with coupled quantum system as the working substance, the effects of
coupling on the cycle is an interesting problem
\cite{Thomas2011PRE}, besides, the thermodynamical relations for the
coupled system and its subsystems are also interesting. Indeed,
previous study shown that in a Otto cycle with coupled quantum
system as its working substance, the total coupled system may absorb
heat from the hot bath and releases heat to the cold bath,  while
the subsystem absorbs heat at the cold bath and releases heat at the
hot bath with a net work done\cite{Thomas2011PRE}.

Motivated by these works, here we study the effects of coupling on
the quantum isobaric process and the quantum Brayton cycle. Two
coupled spins are considered  as the working substance, the
thermodynamical  relations for the total system and its subsystem
are studied and several interesting results are observed. These
observations hold true for a general coupled system as the working
substance. This paper is organized as follows. In
Sec.\ref{Sec:singal}, we first give a brief introduction to the
pressure in  quantum processes and quantum Brayton cycle, then we
examine the Brayton cycle with a single spin in external magnetic
field as the working substance. In Sec.\ref{Sec:coupled}, the
detailed analysis for the quantum isobaric process and Brayton cycle
with coupled system as working substance is presented. Discussions on
the generality of our results and conclusions are given in
Sec.\ref{Sec:con}.

\section{Quantum isobaric process and quantum Brayton cycle
for spin-$1/2$ system in an external magnetic field}
\label{Sec:singal}

In this paper, we use the  definition  of quantum pressure  given in
Ref.\cite{Quan2009PRE} as
\begin{eqnarray}
F=-\sum_np_n\frac{d E_n}{d L},\label{eq:Fdefinition}
\end{eqnarray}
where $L$ is the generalized coordinate of the system. This
definition is deduced  from the quantum version of the first law of
thermodynamics $dU=\sum_nE_ndp_n+\sum_np_ndE_n\equiv \dbar Q+\dbar
W=TdS+\sum_n Y_ndy_n$ and an analogy of classical relation between
the generalized force and the generalized coordinate as
$Y_n=-\frac{\dbar W}{dy_n}$, where $T$ and $S$ refer to temperature
and thermodynamical entropy, respectively, $Y_n$ is the generalized
force and $y_n$ is its conjugated generalized coordinate ($dy_n$ can
be seen as the generalized displacement). We should mention that for
a one dimensional system, the generalized force is the same as
pressure. We first consider a spin-$1/2$ system in an external
magnetic field as the working  substance. The Hamiltonian of the
working substance can be written as $H=\frac12 B\sz$. For the
working substance at thermal equilibrium, choosing the inverse of
the magnetic field as the generalized coordinate $L=1/B$, we can
calculate the generalized force Eq.(\ref{eq:Fdefinition}) as
\begin{eqnarray}
F=-\frac{\tanh(\frac{\beta}{2L})}{2L^2},\label{eq:Fsingle}
\end{eqnarray}
where $\beta=1/kT$, $T$ is the temperature of the system and $k$ is
the Boltzmann constant. Then a quantum isobaric process can be
defined as a  process with  constant force $F$, which can be
realized by controlling  carefully the temperature $T$ (or $\beta$)
and the magnetic field $B$ (or $L$) by Eq.(\ref{eq:Fsingle}). We
should note that the expression for  the pressure  varies from
system to system,  because we do not have a unique equation of
spectrum for quantum system. This was confirmed  in
Sec.\ref{Sec:coupled}.

\begin{figure}
\includegraphics*[width=0.7\columnwidth, bb=140 290 470 560]{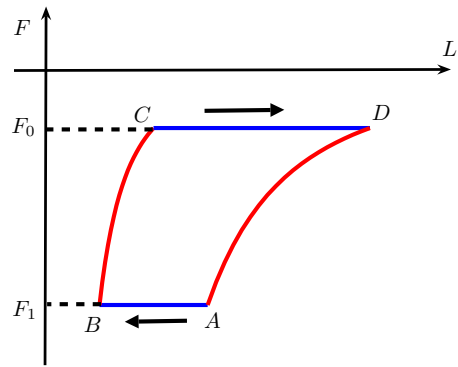}
\caption{(Color online) Force-displacement $F-L$ diagram of a
quantum Brayton cycle, we plot this figure by using the data from
the spin-$\frac 1 2$ as the working substance.}\label{FIG:FL}
\end{figure}

A quantum Brayton cycle is a  generalization of the classical
Brayton cycle \cite{Callenbook,Perrotbook} to quantum case, which
consists of two quantum adiabatic processes and two quantum isobaric
processes (see Fig.\ref{FIG:FL}). Starting from point $A$, the four
stages of a cycle can be depicted as follows: Stage 1: $A\rightarrow
B$ is an isobaric process, in which the generalized force $F_1$
keeps. The system absorbs heat from the environment and some work is
done on the system. In order to ensure that the heat is absorbed
from the environment, we should have $L_B<L_A$. Stage 2:
$B\rightarrow C$ is an adiabatic process, where only some work is
done by the system. Stage 3: $C\rightarrow D$ is almost an inverse
process of stage 1 that the generalized force is replaced by $F_0$.
Stage 4: $D\rightarrow A$ is another adiabatic process. In the
isobaric process the generalized force keeps constant while in the
quantum adiabatic process the entropy of the system is unchanged.
The entropy of the system is
\begin{eqnarray}
S=k[\ln(2\cosh\frac{\beta}{2L})-\frac{\beta}{2L}\tanh\frac{\beta}{2L}].
\end{eqnarray}
It is easy to find that  in the quantum adiabatic process we have
$TL{=}$const (or $\beta/L{=}$const). Based on this fact, we can get
a further equation in quantum  adiabatic process from
Eq.(\ref{eq:Fsingle}) as
\begin{eqnarray}
FL^2=\text{const}.
\end{eqnarray}
The internal energy of the system is
\begin{eqnarray}
U=-\frac1{2L}{\tanh{\frac{\beta}{2L}}}.
\end{eqnarray}
Comparing it with Eq.(\ref{eq:Fsingle}) we find $FL=U$.  These two
relations together yield the basic thermodynamics quantities such as
heat transfer, net work done by the system and efficiency, as
follows,
\begin{eqnarray*}
Q_{AB}&=&\int_{L_A}^{L_B}\sum_nE_ndp_n\\
&=&\left.\sum_np_nE_n\right|_{L_A}^{L_B}-\int_{L_A}^{L_B}\sum_np_n\frac{dE_n}{dL}dL\\
&=&U(L_B)-U(L_A)+\int_{L_A}^{L_B}F_1dL=2F_1(L_B-L_A),\\
Q_{CD}&=&2F_0(L_C-L_D),\\
W&=&Q_{AB}-Q_{CD},\\
\eta &=& 1-\frac{Q_{CD}}{Q_{AB}}=1-\sqrt{\frac{F_0}{F_1}}.
\end{eqnarray*}
We can see from this equation that the system absorbs heat in the
process $A\rightarrow B$ and releases heat in the process
$C\rightarrow D$. The efficiency is consistent to the classical
result $\eta=1-(\frac{F_0}{F_1})^{1-1/\gamma}$, where $\gamma$ is
the adiabatic exponent. In this model, we have shown that in the
quantum adiabatic process $TL{=}$const. Comparing with
$TL^{\gamma-1}{=}$const for the classical adiabatic process, the
adiabatic exponent is $\gamma=2$. These result can be found in the
coupled system as the working substance, as we will show below.

\section{Quantum isobaric process and quantum Brayton cycle
with two coupled spin-$1/2$s as the working substance}
\label{Sec:coupled}

The main purpose of this paper is to study the quantum isobaric
process and quantum Brayton cycle for a coupled system as the
working substance and to consider the effect of coupling on the
cycle. The Hamiltonian for  the working substance under
consideration is
\begin{eqnarray}
H=\frac{B}2(\sz_1+\sz_2)+J(\sigmap_1\sigmam_2+\sigmam_1\sigmap_2),
\label{eq:coupledH}
\end{eqnarray}
where $J$ is the coupling constant between the spins. $J>0$ and
$J<0$ correspond to the antiferromagnetic and the ferromagnetic
case, respectively. Here we only consider the antiferromagnetic
case, i.e., $J>0$. The four eigenvalues and corresponding
eigenstates can be easily obtained as
\begin{eqnarray}
&&E_1=-B,~~~\ket{\psi_1}=\ket{00},\nonumber\\
&&E_2=-J,~~~~\ket{\psi_2}=\frac1{\sqrt{2}}(\ket{01}-\ket{10}),\nonumber\\
&&E_3=J,~~~~~~\ket{\psi_3}=\frac1{\sqrt{2}}(\ket{01}+\ket{10}),\nonumber\\
&&E_4=B,~~~~~\ket{\psi_4}=\ket{11}.\label{eq:eigencoup}
\end{eqnarray}
There are two independent parameters in the Hamiltonian, and hence
we need two generalized coordinates to describe the system. Choosing
$X=1/B$ and $Y=1/J$ as the generalized coordinates, we can define
two generalized forces (or pressures) corresponding to $X$ and $Y$,
respectively, as
\begin{eqnarray}
F_x=-\frac{\sinh\frac{\beta}{X}}{X^2(\cosh\frac{\beta}{X}+\cosh\frac{\beta}{Y})},\label{eq:Fx}\\
F_y=-\frac{\sinh\frac{\beta}{Y}}{Y^2(\cosh\frac{\beta}{X}+\cosh\frac{\beta}{Y})}.\label{eq:Fy}
\end{eqnarray}
The quantum isobaric process for this system means either $F_x$ or
$F_y$ is fixed. As a result, we will discuss these two different
cases separately.  The expression for the entropy is complicated,
hence we do not want to write it here. Noting that the entropy in a
quantum adiabatic process is unchanged, we have
\begin{eqnarray}
XT=\text{const},~~YT=\text{const},~~\frac{X}{Y}=\text{const}.\label{eq:adiabcouple}
\end{eqnarray}
This means that all the spacings of energy-level  for  the working
substance change by the same ratio, this together with  another
restriction we adopted in Sec.\ref{Sec:singal}, i.e., this ratio
equals to the ratio of the temperature of the substance before the
adiabatic process to that after the adiabatic process, guarantee the
reversible cycle. The discussion in this paper focuses exactly on
this reversible quantum Brayton cycle. The internal energy of the
system can be written as
\begin{eqnarray}
U=-\frac{Y\sinh\frac{\beta}{X}+X\sinh\frac{\beta}{Y}}
{XY(\cosh\frac{\beta}{X}+\cosh\frac{\beta}{Y})}=F_xX+F_yY,\label{eq:Ucoupled}
\end{eqnarray}
which will be used in the following discussions. We should also note
that in the following discussion, we consider the relation between
the total system and subsystem in the cycle. We will denote the
symbol $F_{\text{loc}}$ or the concept of local force to  the force
corresponding to the local magnetic field $B$ for one of the
subsystems.

\subsection{Fixed $F_x$}\label{sec:Fx}
We first consider the relation between the composite system and its
subsystems in quantum isobaric process when the generalized force
$F_x$ is fixed. In this process, the generalized coordinate and the
temperature are controlled according to Eq.(\ref{eq:Fx}) such that
$F_x$ is a constant with the generalized coordinate $Y$ as  another
constant. We will prove that both of the subsystems undergo a
quantum isobaric process when the generalized force $F_x$ is fixed
and the generalized force for the subsystems equals to $\frac12F_x$.
As a consequence, the cycle for the subsystem is also a quantum
Brayton cycle.

\emph{Proof}: the state of the composite system is in a thermal
equilibrium state which can be easily calculated as
\begin{eqnarray}
\rho(T){=}\sum_n p_n \ext{\psi_n}{\psi_n}{=}\left(
                                          \begin{array}{cccc}
                                            p_4 & 0 & 0 & 0 \\
                                            0 & \frac12(p_2{+}p_3) & \frac12(p_3{-}p_2) & 0 \\
                                            0 & \frac12(p_3{-}p_2) & \frac12(p_2{+}p_3) & 0 \\
                                            0 & 0 & 0 & p_1 \\
                                          \end{array}
                                        \right),
\end{eqnarray}
where $p_n=\frac1{\mathcal Z}\exp(-\frac{E_n}{kT})$, and $T$ is the
temperature of the composite system. ${\mathcal
Z}=\sum_n\exp(-\frac{E_n}{kT})$ is the partition function of the
system. For the reduced system, we can get the reduced state by
tracing out its partner as
\begin{eqnarray}
\rho_A=\rho_B=\left(
                \begin{array}{cc}
                  p_4+\frac12(p_2+p_3) & 0 \\
                  0 & p_1+\frac12(p_2+p_3) \\
                \end{array}
              \right).
\end{eqnarray}
This state can be seen as an equilibrium state with a local effective temperature
$\beta_{\text {loc}}$ (or $T_{\text {loc}}$) as
\begin{eqnarray}
\beta_{\text{loc}}=X\ln \frac{p_1+\frac12(p_2+p_3)}{p_4+\frac12(p_2+p_3)}.
\end{eqnarray}
According to the definition of generalized force for a single spin
in Sec.\ref{Sec:singal}, we arrive at
\begin{eqnarray}
F_{\text{loc}}=-\frac1{2X^2}\tanh(\frac{\beta_{\text{loc}}}{2X}).
\end{eqnarray}
By virtue of the formula $\tanh(\frac12\ln x)=\frac{x-1}{x+1}$ and the definition for
$F_x$ (Eq.(\ref{eq:Fx})) we have
\begin{eqnarray}
F_{\text{loc}}=-\frac{\sinh\frac{\beta}{X}}{2X^2(\cosh\frac{\beta}{X}+\cosh\frac{\beta}{Y})}
=\frac12F_x.
\end{eqnarray}
This result shows that if the composite system keeps the generalized
force $F_x$ unchanged, the generalized force for subsystems is also
a constant  and it equals to $\frac12F_x$.

\begin{figure}
\includegraphics*[width=0.46\columnwidth, bb=95 255 470 576]{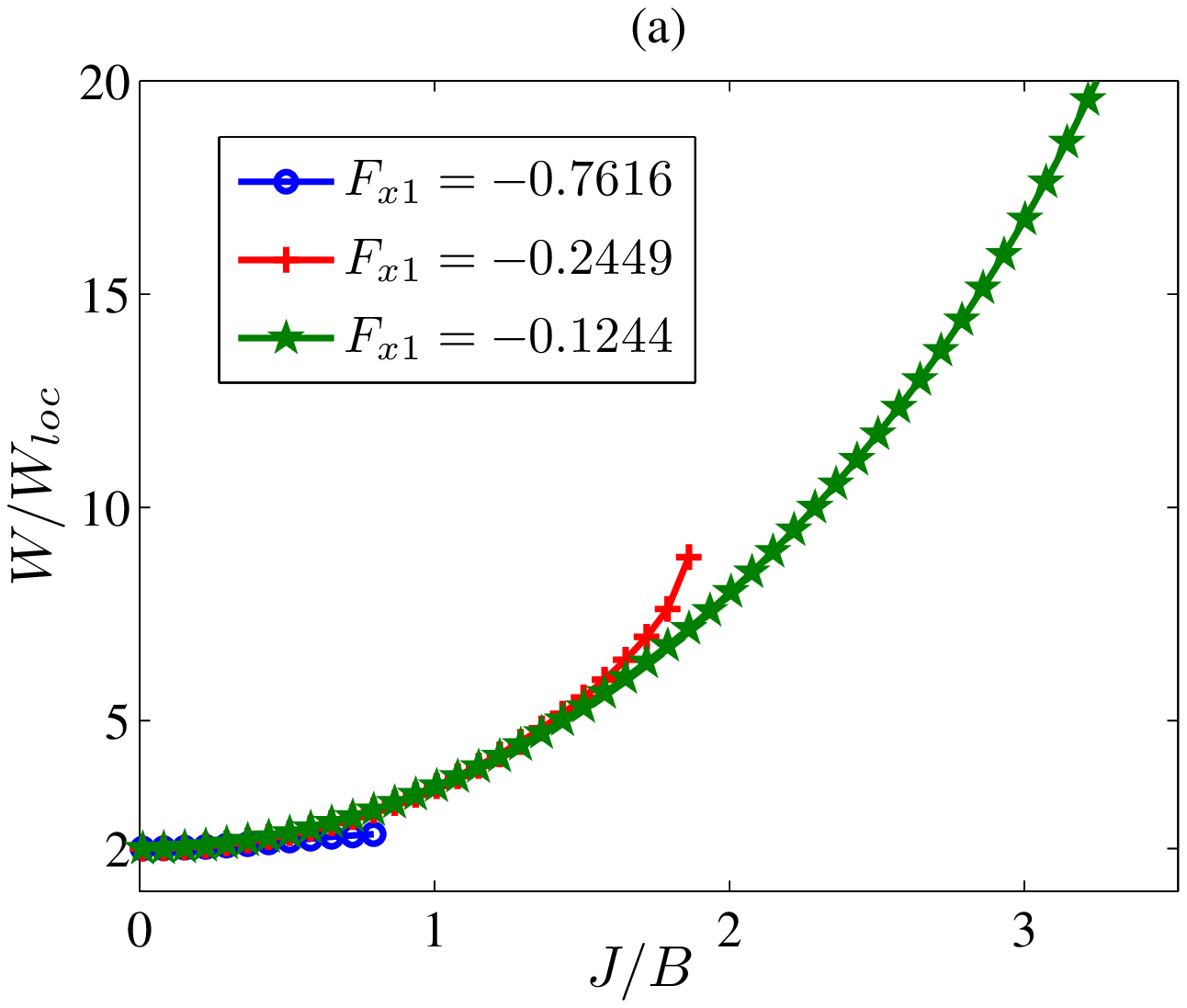}
\includegraphics*[width=0.46\columnwidth, bb=95 255 470 576]{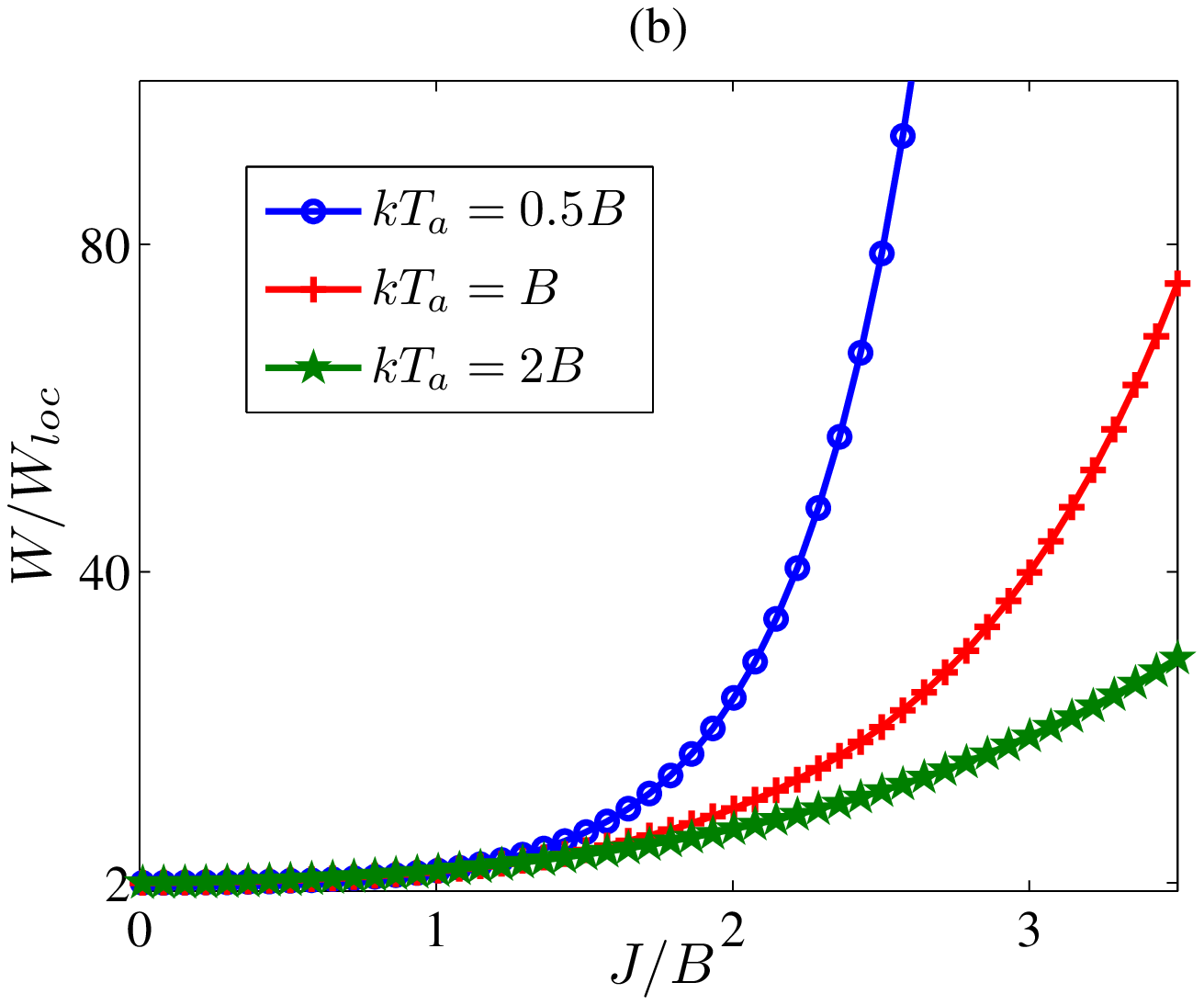}
\caption{(Color online) Efficiency and net work done by the coupled
system as a function of $J/B$ at point $A$ when (a) initial pressure
$F_{x1}$ (b) initial temperature $kT_a$ is fixed. Other parameters
in the figure are $F_{x0}=\frac14F_{x1}$,
$\frac{X_A}{X_B}=\frac{X_D}{X_C}=3$.}\label{FIG:W_JFxfix}
\end{figure}

In the following, we consider the quantum Brayton cycle based on the
coupled system when $F_x$ is fixed. Similar to the case for a single
spin, we decrease $X$ in stage 1 and adjust the temperature $T$, so
that $F_{x1}$ keeps unchanged. Moreover, the parameters $Y_1$, which
is the inverse of the coupling constant, is also unchanged in this
stage. In the adiabatic process, Eq.(\ref{eq:adiabcouple}) should be
satisfied so that the cycle is reversible. After this stage, the
parameter becomes $Y_0$ and in stage 3 it keeps as a constant. The
force in stage 3 is $F_{x0}$. The heat absorbed by the system during
the stage 1 is
\begin{eqnarray}
&&Q_{AB}{=}\int_{X_A}^{X_B}\sum_nE_ndp_n
{=}U(X_B){-}U(X_A){+}\int_{X_A}^{X_B}F_{x1}dX\nonumber\\
&&~~~~=2F_{x1}(X_B-X_A)+(F_{yB}-F_{yA})Y_1.
\end{eqnarray}
We can see from the above expression that $Q_{AB}$ depends not only on the
force $F_x$ but also on the force $F_y$ and the coupling constant. This means
that the interaction between the two spins has a strong effect on the cycle.
Similarly, we can calculate the heat released to environment in stage 3 as
\begin{eqnarray}
Q_{CD}=2F_{x0}(X_C-X_D)+(F_{yC}-F_{yD})Y_0.
\end{eqnarray}
Hence the efficiency of the cycle can be expressed as
\begin{eqnarray}
\eta_x=1-\frac{Q_{CD}}{Q_{AB}}=1-\sqrt{\frac{F_{x0}}{F_{x1}}}. \label{eq:effcouple}
\end{eqnarray}
The detailed derivation about
the above equation is given in appendix \ref{sec:eta}.
This result is the same to the single spin case. After the same procedure
as given in Sec.\ref{Sec:singal}, we can get the efficiency for the subsystem
as
\begin{eqnarray}
\eta_{\text{loc}}=1-\sqrt{\frac{F_{\text{loc}0}}{F_{\text{loc}1}}}
=1-\sqrt{\frac{F_{x0}}{F_{x1}}},
\end{eqnarray}
This equation shows that the efficiency for the composite system as
working substance is the same to the one for its subsystem. The net work done
by the composite system is
\begin{eqnarray}
&&W=[2F_{x1}(X_B-X_A)-2F_{x0}(X_C-X_D)]\nonumber\\
&&~~~~~~~~~~~~+[(F_{yB}-F_{yA})Y_1-(F_{yC}-F_{yD})Y_0].
\end{eqnarray}
The first $[\cdot]$ on the right hand side can be seen as the sum of
the work done by the two subsystem while the second $[\cdot]$ comes
from the effects of interaction between the two subsystems. Due to
the positivity of this term, we have $W>2W_{\text{loc}}$, i.e., the
total work performed is larger than the sum of work obtained from
the two spins locally. Numerical examples for this result is shown
in Fig.\ref{FIG:W_JFxfix}. From the figure we claim  that the
coupling can increase the net work done by the system during a cycle
although the efficiency is not  improved. When $J=0$,
$W/W_{\text{loc}}=2$ in all cases as expected. Another point to be
explained is that in Fig.\ref{FIG:W_JFxfix}, some lines are crossed.
The reason is  that when the absolute value of $F_{x1}$ is larger,
the possible region for the coupling constant is smaller.

\subsection{Fixed $F_y$}
Similar to the discussion given above, an isobaric process that
$F_y$ is fixed means one should carefully control the coordinate $Y$
and temperature $T$ such that $F_y$ is a constant with $X$ as
another constant. It can be easily verified that during this
process, the pressure for the subsystem is not fixed, i.e., the
process for the subsystem is not a quantum isobaric process.
However, when we control the coordinate $Y$ for the composite
system, the local Hamiltonian for the subsystem does not vary and
the energy levels for the subsystem   keep as constants. As a
result, the subsystems undergo quantum isochoric  process when $F_y$
is fixed, and a quantum Brayton cycle based on $F_y$ for the
composite system results in a quantum Otto cycle for the subsystem.

\begin{figure}
\includegraphics*[width=0.9\columnwidth]{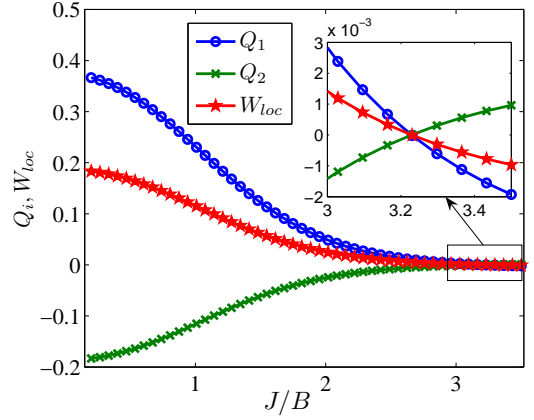}
\caption{(Color online) The local heat exchange $Q_1$ and $Q_2$ and
the local work $W_{\text{loc}}$ (in units of $B$) as a function of
$J/B$. Here the initial temperature $kT_a=0.5B$. Other parameters
are $F_{y0}=\frac14F_{y1}$,
$\frac{Y_A}{Y_B}=\frac{Y_D}{Y_C}=3$.}\label{FIG:WlocFy}
\end{figure}
After some calculations which are similar to Sec.\ref{sec:Fx}, we
obtain the efficiency of the cycle for the composite system as
\begin{eqnarray}
\eta_y=1-\sqrt{\frac{F_{y0}}{F_{y1}}}.\label{eq:effcoupley}
\end{eqnarray}
The efficiency for the subsystem can be obtained according to the efficiency of
quantum Otto cycle as
\begin{eqnarray}
\eta_{\text{loc}}=1-\frac{X_1}{X_0}=1-\sqrt{\frac{F_{y0}}{F_{y1}}}.
\end{eqnarray}
This result tells us that the efficiency for the composite system
during a quantum Brayton cycle with fixed $F_y$ equals to the one
for the subsystem during a quantum Otto cycle, conditioned on  that
the subsystem is a Otto heat engine. We should note that when
initial temperature is small enough and the coupling constant is
large enough, the work done by the subsystem can be negative. In
this case, the subsystem is a refrigerator while the total system is
a heat engine. Numerical examples are given in Fig.\ref{FIG:WlocFy}.
In this figure we show  the heat exchange $Q_1$ in stage 1 and $Q_2$
in stage 3 for the subsystem. Here $Q_i>0$ and $Q_i<0$ ($i=1,2$)
correspond to absorption and release of heat, $W_{\text{loc}}$ is
the local work. From the figure we see that when initial $J/B$ at
point $A$ is larger than a certain value, $Q_1<0$, $Q_2>0$ and
$W_{\text{loc}}<0$. This indicates that the subsystem is a
refrigerator, it releases heat when the composite system absorbs
heat in stage 1 and goes opposite  in stage 3. The condition for
such a phenomenon is that $p_e^B<p_e^A$, where
$p_e^i=p_4^i+\frac12(p_2^i+p_3^i)$ is the probability for the
subsystem on its excited state at point $i (i=A,B)$. The  feature
comes from the strange property for the eigenstate $\ket{\psi_3}$ of
the system. In the strong coupling regime, the ground state of the
system is $\ket{\psi_3}$. However, the reduced state for this state
is $\rho_{A(B)}=\frac12I_2$, which can be seen as an equilibrium
state at a local effective temperature $T_{\text{loc}}=\infty$. Here
$I_2$ is a $2\times2$ unit matrix. Hence in stage 1,  when the
global temperature is low enough and the coupling constant is large
enough, the global temperature increases while the local temperature
decreases. Hence the subsystem releases heat in stage 1 and
inversely in stage 3. As a result the local cycle is a refrigeration
cycle.

\section{Discussions and Conclusions}\label{Sec:con}
Before concluding, we briefly  discuss the generality of our
results. The conclusion holds true for  two spin-$\frac 1 2 $s with
general couplings described by the following Hamiltonian,
$H=\frac{J}2[(1+\gamma)\sx_1\sx_2+(1-\gamma)\sy_1\sy_2]+J\Delta\sz_1\sz_2
+\frac{B}2(\sz_1+\sz_2)$, where $\gamma$ and $\Delta$ are anisotropy
parameters. The four eigenvalues and corresponding eigenstates for
this Hamiltonian are $E_1=J\Delta-\sqrt{B^2+J^2\gamma^2},\ket{\psi_1}
=\cos\theta\ket{11}-\sin\theta\ket{00},E_2=-J(1+\Delta),\ket{\psi_2}
=\frac1{\sqrt{2}}(\ket{01}-\ket{10}),E_3=J(1-\Delta),\ket{\psi_3}
=\frac1{\sqrt{2}}(\ket{01}+\ket{10}),E_4=J\Delta+\sqrt{B^2+J^2\gamma^2},\ket{\psi_4}
=\sin\theta\ket{11}+\cos\theta\ket{00}$, where
$\tan\theta=\frac{B+\sqrt{B^2+J^2\gamma^2}}{J\gamma}$.
 If  $\gamma=\Delta=0$, the model returns to the
Hamiltonian given in Eq.(\ref{eq:coupledH}). For the system with
general couplings, we still choose $X=\frac1B$ and $Y=\frac1J$ as
the generalized coordinates and define the generalized forces $F_x$
and $F_y$ similar to our earlier discussions in this paper. For the
quantum Brayton cycle characterized by the generalized force $F_x$,
we can prove in the same manner that the cycle for the subsystem is
also a quantum Brayton cycle and the efficiencies for the subsystem
and total system are equal.  However, for the quantum Brayton cycle
with fixed $F_y$, the cycle for the subsystem is a Otto cycle. The
situation that  the total system is a heat engine while the
subsystem operators as a refrigerator can  happen when
$p_e^B<p_e^A$, where $p_e$ is the population of the excited state
at point $i(i=A,B)$. This
is similar to the discussion in the last section. For the two spins
with generalized couplings, it is involved to give an analytical
result, the following analysis would help understanding the
prediction  in such systems.

We now focus on the result that the total system behaves like a heat
engine but the subsystem behaves like a refrigerator. The essence of
this result is that in one of the four stages of the cycle the total
system absorbs heat while the subsystem releases heat. A similar
result was observed in Ref.\cite{Thomas2011PRE}, where a coupled
Otto cycles was considered. There are four basic thermodynamical
processes in quantum thermodynamics. It is obviously that this
feature can not happen in a quantum adiabatic process. Based on our
results and the results in Ref.\cite{Thomas2011PRE}, we know that it
can happen in quantum isobaric process and quantum isochoric
process. The only process that has not been discussed to date is the
quantum isothermal process for coupled systems in which  the
situation is more complicated because the process for the subsystem
might be not any one  of the four basic thermodynamical processes.
Hence the heat transfer for the subsystem is difficult to handle.
Here we only give an extreme example by considering  the system
given in Eq.(\ref{eq:coupledH}) undergoes a quantum isothermal
process with absolute zero temperature $T=0$. In this process, the
magnetic field $B$ decreases slowly so that the system is always in
the ground state. Initially, $B>J$, the ground state is
$\ket{\psi_1}=\ket{00}$. As a result, the subsystem are also in the
ground state (effective temperature $T_{\text{eff}}=0$). After some
time, when $B<J$, $\ket{\psi_2}$ becomes the ground state of the
total system. Hence, the state for the subsystem becomes
$\rho_A=\rho_B=\frac12I_2$, a state with the effective temperature
$T_{\text{eff}}=\infty$. In this process, the generalized
coordinate, generalized force, effective temperature and entropy for
the subsystem are all changed. As a result, none of the four basic
thermodynamical processes match the behavior of the subsystem. The
case in which the heat transfer is different between the total
system and subsystem may also happen under certain condition in an
isothermal process. This can be understood as follows. From the
energy spectrum of the composite system given in
Eq.(\ref{eq:eigencoup}) we know that the entropy of the system
$S=S(B,J)=-k\sum_np_n\ln p_n$ is symmetric  about $B$ and $J$, i.e.,
this expression does not change when we exchange the two variables
$B$ and $J$. Then we take a total differential for $S$ as
$dS=\left(\frac{\partial S}{\partial B}\right)dB+
\left(\frac{\partial S}{\partial J}\right)dJ$. Initially we set
$B=J$. Due to the symmetry of the entropy, we have
$\left(\frac{\partial S}{\partial B}\right)=\left(\frac{\partial
S}{\partial J}\right)<0$. Now we consider two processes as follows:
(i) $B\rightarrow B+dB$; (ii) $J\rightarrow J+dJ$. Here $dB$ and
$dJ$ denote infinitesimal  increment for magnetic field and coupling
constant, respectively. We assume $dB>0$ and $dJ>0$. It can be
easily proved that $dS<0$ for these two processes. Hence in these
two isothermal processes, $\dbarit Q=TdS<0$, i.e., the total system
releases heat to the heat bath in both cases. However, the heat
transfer are different for  the subsystem in these two processes.
This can be seen obviously in the limit $\beta\rightarrow\infty$ (or
$T\rightarrow0$, i.e. the low temperature behavior). With  this
limitation and initial condition $B=J$, we can obtain the state of
the subsystem as
$\rho_{Ai}=\rho_{Bi}=\frac14\ext{\uparrow}{\uparrow}
+\frac34\ext{\downarrow}{\downarrow}$ (In fact, when $T\neq0$, this
is not the exact state for the subsystem, however when the
temperature is low enough, the probability of the subsystem in this
state is higher that 0.99, hence we analyze the heat transfer
according to this approximate state). Then we consider the following
two processes separately. (i) $B\rightarrow B+dB$. In this case,
$\ket{\psi_1}$ is the ground state of the total system. As a result,
$\rho_{Af}=\rho_{Bf}=\ext{\downarrow}{\downarrow}$. During this
process, the internal energy of the subsystem decreases. The work
done in this process is infinitesimal. Hence the subsystem releases
heat. (ii) $J\rightarrow J+dJ$. $\ket{\psi_2}$ becomes the ground
state of the total system, then $\rho_{Af}=\rho_{Bf}=\frac12I_2$.
Hence the subsystem absorbs heat. We can see in the case (ii) that
the heat transfer is different between the total system and the
subsystem in a quantum isothermal process. Based on the discussion
above, we conclude that the thermodynamical cycle which consists of
the quantum adiabatic process and one or two of other three
processes (for example, Carnot cycle and Diesel cycle) can have the
similar property that the heat flow of the subsystem and total
system are in the opposite directions, namely, the total system
absorbs heat while the subsystem releases heat.

In summary, with coupled spins as working substance, the quantum
isobaric process and the quantum Brayton cycle have been studied in this
paper. The concept of force or pressure in quantum system is deduced
from the quantum version of the first law of thermodynamics and an
analogy of classical relation between generalized force and
generalized coordinate \cite{Quan2009PRE}. The quantum Brayton cycle
consists of  two quantum adiabatic processes and two quantum
isobaric processes. There are two generalized coordinates in our
system, i.e., the local external magnetic field and the coupling
constant, therefore we can define two pressures respectively and
construct two types of quantum Brayton cycle. We find that in
quantum Brayton cycle based on the pressure corresponding to the
external field, the subsystem undergo a quantum Brayton cycle with
pressure half of the total system, while in the cycle based on the
force conjugated to the coupling strength, the subsystem experiences
a quantum Otto cycle. The efficiency for the coupled system in the
two cycles are equal to that for the subsystem, which is the same as
 the classical result, but the net work done by the total system
are usually more than the sum of works done by the subsystems.
Moreover, when the initial temperature and initial coupling strength
are chosen properly, an interesting phenomenon can be observed in
the Brayton cycle based on the force corresponding to coupling
constant, i.e.,  the total system performed as a heat engine while
the subsystems serve as a refrigerator. The essence for this
interesting result is that the heat flows to different directions in
the subsystems and in the whole system. This can happen in quantum
isochoric process and quantum isothermal process, which is a
reminiscence of  the non-locality of quantum system in
thermodynamics.

\ \ \\
This work is supported by NSF of China under Grant Nos. 11105064, 10905007,
and 11175032.

\appendix
\section{The derivation of $\eta$ for the coupled system when $F_x$ is fixed.}\label{sec:eta}
First, from expression of $F_x$ in Eq.(\ref{eq:Fx}) and the condition for adiabatic
process in Eq.(\ref{eq:adiabcouple}), we have
\begin{eqnarray}
F_xX^2=\text{const}
\end{eqnarray}
As a result, one can construct the relation during the two adiabatic processes as
\begin{eqnarray}
\frac{F_{x0}}{F_{x1}}=\frac{X_B^2}{X_C^2}=\frac{X_A^2}{X_D^2}.\label{eq:A1}
\end{eqnarray}
In the similar manner, for the force $F_y$ we obtain
\begin{eqnarray}
F_yY^2=\text{const}
\end{eqnarray}
and
\begin{eqnarray}
\frac{Y_1^2}{Y_0^2}=\frac{F_{yC}}{F_{yB}}=\frac{F_{yD}}{F_{yA}}.\label{eq:A2}
\end{eqnarray}
Moreover, based on $\frac{X}{Y}=\text{const}$ in the adiabatic process, we have
\begin{eqnarray}
\frac{Y_1}{Y_0}=\frac{X_A}{X_D}=\frac{X_B}{X_C}\label{eq:A3}
\end{eqnarray}
Combining Eqs.(\ref{eq:A1}), (\ref{eq:A2}) and (\ref{eq:A3}), we obtain a complete
ratio relation for the whole cycle as
\begin{eqnarray}
\frac{Y_1^2}{Y_0^2}=\frac{X_A^2}{X_D^2}=\frac{X_B^2}{X_C^2}=
\frac{F_{x0}}{F_{x1}}=\frac{F_{yC}}{F_{yB}}=\frac{F_{yD}}{F_{yA}}.
\end{eqnarray}
Using this equation we can obtain Eq.(\ref{eq:effcouple}).
In a similar way Eq.(\ref{eq:effcoupley}) can be obtained.

\end{CJK*}


\begin{references}
\bibitem{Born} M. Born and V. Fork. Z. Phys. 51, 165 (1928).

\bibitem{Messiah} A. Messiah, \emph{Quantum Mechanics} (Dover, New York, 1999).

\bibitem{Kieu2004PRL} T. D. Kieu, \PRL{93}{140403}{2004}.

\bibitem{Kieu2006EPJD} T. D. Kieu, \EPJD{39}{115}{2006}; quan-ph/0311157.

\bibitem{Quan2005PRE} H. T. Quan. P. Zhang, and C. P. Sun,
\PRE{72}{056110}{2005}.
\bibitem{Quan2007PRE} H. T. Quan, Y. X. Liu, C. P. Sun, and F. Nori,
\PRE{76}{031105}{2007}.

\bibitem{Quan2009PRE} H. T. Quan, \PRE{79}{041129}{2009}.

\bibitem{Callenbook} H. B. Callen, \emph{Thermodynamics and an Introduction to Themostatistics},
2nd ed. (Wiley, New York, 1985); C. Kittle and H. Kroemer, \emph{Thermal Physics}, 2nd ed.
(Freeman, San Francisco, 1980).

\bibitem{Perrotbook} P. Perrot, \emph{A to Z of Thermodynamics}
(Oxford University Press, Oxford, 1998).




\bibitem{Fialko2012PRL} O. Fialko and D. W. Hallwood, \PRL{108}{085303}{2012}.

\bibitem{He2002PRE} J. Z. He, J. C. Chen, and B. Hua,
\PRE{65}{036145}{2002}.


\bibitem{FeldmannPRE}T. Feldmann and R. Kosloff, \PRE{61}{4774}{2000};
\PRE{68}{016101}{2003}; \PRE{70}{046110}{2004}.

\bibitem{Wu2006JCP} F. Wu, L. G. Chen, S. Wu, F. R. Sun, C. Wu,
\JCP{124}{214702}{2006}.


\bibitem{Lin2003PRE} B. H. Lin and J. C. Chen, \PRE{67}{046105}{2003}.
\bibitem{Li2007JPA}S. Li, H. Wang, Y. D. Sun, and X. X. Yi, \JPA{40}{8655}{2007}.






\bibitem{Scully2003Science} M. O. Scully, M. S. Zubairy, G. S. Agarwal,
and H. Walther, \Science{299}{862}{2003}.

\bibitem{WangPRE} J. H. Wang, H. Z. He, and X. He, \PRE{84}{041127}{2011};
J. H. Wang, H. Z. He, and Z. Q. Wu, \PRE{85}{031145}{2012}; J. H. Wang, Z. Q. Wu,
and J. Z. He, \PRE{85}{041148}{2012}; J. H. Wang and J. Z. He,
\JAP{111}{043505}{2012}.


\bibitem{Huangsubmit}X. L. Huang, Tao Wang, and X. X. Yi, \PRE{86}{051105}{2012}.

\bibitem{ScullyPRL} M. O. Scully, \PRL{87}{220601}{2001}; \PRL{88}{050602}{2002}.

\bibitem{Quan2006} H. T. Quan, Y. D. Wang, Y. X. Liu, C. P. Sun, and F. Nori,
\PRL{97}{180402}{2006}; H. T. Quan. P. Zhang, and C. P. Sun,
\PRE{73}{036122}{2006}.

\bibitem{Zhang2007PRA} T. Zhang, W. T. Liu, P. X. Chen, and C. Z. Li,
\PRA{75}{062102}{2007}.

\bibitem{Wang2009PRE} H. Wang, S. Q. Liu, and J. Z. He, \PRE{79}{041113}{2009}.

\bibitem{Zhang2008EPJD} G. F. Zhang, \EPJD{49}{123}{2008}.

\bibitem{Thomas2011PRE} G. Thomas and R. S. Johal, \PRE{83}{031135}{2011}.



%


\end{references}
\end{document}